# Total and Partial Pressure Measurement

*K. Jousten*
Physikalisch-Technische Bundesanstalt, Berlin, Germany

**Abstract**
Total pressure gauges are used to measure the level of vacuum in a system independent of the gas species. Partial pressure analysers measure the occurrence of each gas species separately. This report explains the methodology of measurement, summarizes the measurement principles of total pressure vacuum gauges, and updates the standardization work for their calibration. On the matter of partial pressure measurement, this report explains the mass filtering by the quadrupole mass spectrometers and outlines the measurement problems with these instruments.

## 1 Introduction

In accelerators, total pressure gauges are mainly used to switch between pumps at the suitable pressure levels and to characterize the functionality of the vacuum system, while partial pressure analysers are mainly used to measure specific partial pressures to detect leaks or problematic outgassing or for desorbing gas species.

At the two CERN accelerator schools, (CAS) `Vacuum Technology' in 1999 and 'Vacuum in accelerators' in 2006, broad reviews of gauges for fine and high vacuum [1][2] and ultrahigh vacuum gauges [3][4] were given. Since the technology of vacuum gauges has not significantly changed in the past two decades, the reader is referred to these publications for the history, measurement principles, and designs of vacuum gauges. This report will mainly complement these publications with metrological issues and will give an update on the standards developed for vacuum gauge calibration and measurement. Partial pressure measurement, however, was covered only in a short report [5] of the CAS in 1999 and will be described in more detail here.

## 2 The Measurement of Vacuum and Some Important Metrological Terms

Vacuum is measured by the physical quantity of pressure. The value of a physical quantity $Q$ is expressed as product of a number $\{Q\}$ and a unit $[Q]$.

$$Q = \{Q\} \cdot [Q]. \tag{1}$$

This equation shows that each measurement process is a comparison between a physical quantity and a unit. The unit is agreed on by convention and is provided by a calibration. In the framework of the international system of units (SI) the pascal, the unit for pressure, may be defined as

$$1\,[Pa] = [N][m]^{-2} = [kg][m]^{-1}[s]^{-2}. \tag{2}$$

It is, 1 [Pa] = 0.01 [mbar]. The 'mbar' was not defined in the SI of 2018, but it is still commonly used and is easy to convert to the pascal. The units 'Torr' and 'micron', however, are related to the height of a mercury column and not to the basic SI units and are obsolete.

It is a matter of discussion whether the quantity of pressure is a reasonable quantity to use to measure vacuum or not [6]. The advantage of the unit of pressure is that there is one common scale, from the ultrahigh vacuum up to the highest technically available high-pressure states of a gas as occur in an engine. The pressure of a gas system, however, is always related to the temperature of the same system. This relation is expressed by the ideal gas law or the real gas equation. In vacuum applications, however, in particular in vacuum systems for accelerators, the temperature is of minor importance, since the main purpose of the vacuum is to achieve a long mean free path of the accelerated particles. The mean free path, however, solely depends on the gas density and not on the temperature. Thus, alternatively, a volume density such as the 'number of molecules per m³' or 'number of moles per m³' could be chosen as a more suitable quantity to measure vacuum. Also, in industrial applications, e.g., for plasmas and for coating applications, the gas density is the crucial quantity which has to be controlled. It remains to be seen if future generations will choose a different quantity than pressure for measuring vacuum, in particular, when new optical methods to measure gas density are established [7].

Any unit [$Q$] must be realized by a fundamental standard. In the case of the pascal this fundamental standard applies a well-known physical law and gives the most direct link to the underlying SI units kg, m, and s. Due to the wide scale of pressure, several fundamental standards for different regimes of vacuum are in operation and these apply different physical laws to define pressure: the equation for pressure as the force $F$ acting on area $A$

$$p = \frac{F}{A} \tag{3}$$

is applied by the fundamental standards mercury column and piston gauge, the ideal gas law by the static expansion systems, and the law of continuity by the continuous expansion standards. An overview of these fundamental methods and standards is given in Ref. [8]. Except for the mercury column, the fundamental standards are pressure generators, which means that in the standards, a well-known pressure is generated which can be compared with the reading of a vacuum gauge under calibration.

These fundamental standards for vacuum are realized by several National Metrological Institutes (NMI), e.g., the Physikalisch-Technische Bundesanstalt (PTB) in Germany and the National Institute of Standards and Technology in the United States. They cover the range from $10^{-9}$ Pa up to 100 kPa (atmospheric pressure).

Each value of a generated or measured pressure of a fundamental standard is associated with an uncertainty. This uncertainty gives the possible range within which the value may not reflect the true value defined by the SI units. This uncertainty cannot be reduced by any instrument that is calibrated with such a standard and is therefore a kind of lower limit for the measurement uncertainty of any vacuum gauge. Figure 28 in Ref. [2] gives a rough idea of the relative uncertainties of the fundamental standards.

When vacuum gauges are calibrated on a fundamental standard the uncertainty of the calibration consists of the one from the fundamental standard plus the uncertainties due to the gauge under calibration such as resolution, repeatability etc. A gauge of good metrological quality that is calibrated on a fundamental primary standard is called a secondary standard. When using a secondary standard to calibrate other gauges (e.g., a working standard), further uncertainties have to be considered. Very important in the case of vacuum gauges is the long-term instability which dominates the uncertainty budget in most cases. However, other influences, such as a temperature dependence of the reading of a gauge, may also be important. In this way, with any step of a calibration, the uncertainty increases (Fig. 3 in Ref. [2]).

At this point we want to introduce some important metrological terms as defined by the International Vocabulary of Metrological Terms [9]. An *error* or a *deviation of reading* is a real deviation from a true value defined by the SI units (**Fig. 1**). This must not be mixed with the term uncertainty which specifies the expected possible range of values around the true SI value, as mentioned above (**Fig. 1**). If the value of a gauge is not correct, but the uncertainty around the value includes the true SI value, everything is fine and it is said that the gauge does not have a significant error or deviation.

Manufacturers of vacuum gauges usually specify a *measurement uncertainty* of the gauge in a specification sheet or catalogue. This value must include the scatter of the batch and the measurement uncertainty of an individual gauge (**Fig. 1**). Thus, if a gauge is individually calibrated, its error can be corrected and the uncertainty of the batch can be reduced. It is also important to distinguish between *measurement range* and the *range of indication* of a gauge (**Error! Not a valid bookmark self-reference.**). The measurement range is always equal to or smaller than the indication range. For the measurement range an uncertainty can be given, while this is not always true for the indicated range. Here, the error of a reading is so high (a correction factor of more than 2 or 10 is not unusual for a vacuum gauge) that it does not make sense to specify a measurement uncertainty.

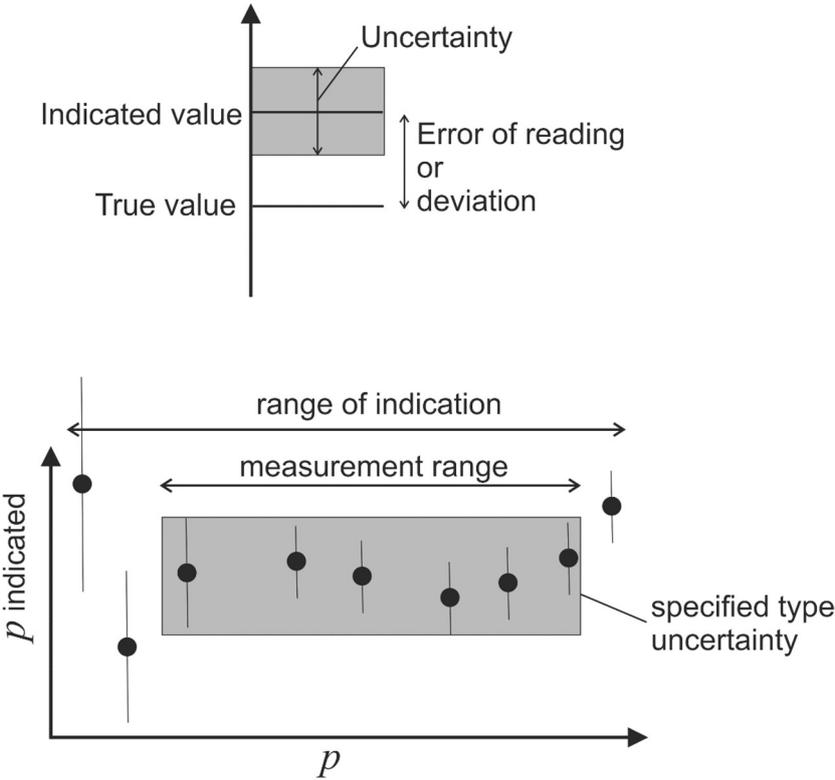

**Fig. 1:** Illustration of the metrological terms error or deviation, uncertainty, measurement uncertainty of a type of gauge, measurement range, and range of indication. The bars around the dots show the uncertainty of an individual gauge reading.

*Accuracy* is defined as the closeness of agreement between a measured quantity value and a true quantity value of a measurement. The concept 'measurement accuracy' is not a quantity and is not given a numerical value. A measurement is said to be more accurate when it offers a smaller measurement error or uncertainty.

# 3 Total Pressure Vacuum Gauges and their Calibration

## 3.1 Measurement principles of total pressure vacuum gauges

**Fig. 2** gives an overview of the ranges and the measurement principles of total pressure vacuum gauges. At the highest vacuum pressures, from atmospheric pressure down to about 100 Pa, the measurement of the exerted force of a gas pressure dominates the measurement principle. The force is exerted on a liquid (oil or mercury) in a U-tube, on a piston (piston gauges), a curved tube (mechanical Bourdon gauges), an oscillating bar dampening its amplitude or changing its frequency (resonance gauges), or a diaphragm (piezo, membrane, capacitance gauges). Capacitance diaphragm gauges (CDGs) are so sensitive that their lower measurement range limit is significantly lower than the mentioned 100 Pa, at about 1 mPa.

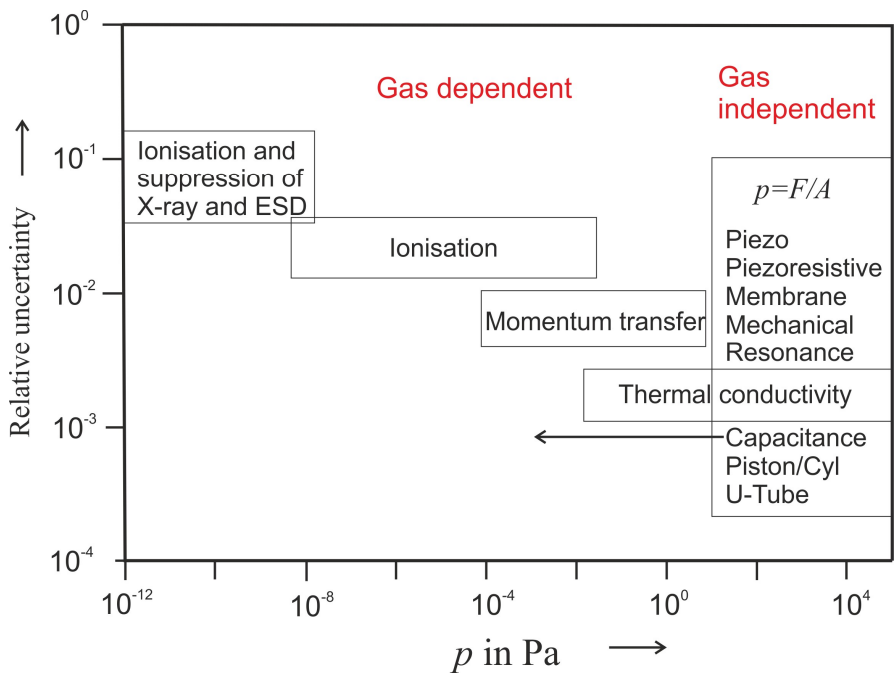

**Fig. 2:** Overview of the ranges and measurement principles of total pressure vacuum gauges. The values of the uncertainties are very approximate and should not be taken as exact values.

When the force exerted by the gas pressure is used, the measurement principle has the significant advantage that it is not sensitive to the type of gas used. This means that it is sufficient to calibrate the gauge using only one gas species.

This is different for thermal conductivity gauges, such as Piranis or thermocouples, which measure the thermal conductivity of the surrounding gas. In the molecular regime, where the molecules travel freely between the heated element and the envelope, the thermal conductivity depends on the number of degrees of freedom of the molecules of the gas species, on its molecular mass via the mean thermal velocity, and on the energy accommodation factor of the molecules on the heated surface [10]. In the viscous flow regime, the accommodation does not play a role, but, in addition to the other mentioned parameters, the mean free path and the ratio of specific heat at constant pressure to that at constant volume do impact on the thermal conductivity. The energy accommodation factor is not predictable and a thermal conductivity gauge must be calibrated for each gas species.

The measurement principle of spinning rotor gauges (SRGs) is the drag force acting on a rotating sphere by momentum transfer which, in the molecular regime important for this gauge, is proportional to the impingement rate of gas molecules. The drag force depends, via the mean thermal velocity, on the molecular mass of the gas molecules and on the momentum accommodation factor of the molecules on the rotor surface. The momentum accommodation factor is predictable, with an uncertainty of about 5%. If a higher accuracy is required, the SRG must be calibrated for each gas species.

In ionization gauges the number of ions generated by a beam of constant current energetic electrons is measured. This number clearly depends on the ionization probability of the molecules by the electrons. The measured signal is also influenced by the secondary yield of the ions hitting the collector, space charge effects, and minor effects such as the charge exchange between molecules. All these additional effects are gas specific and not predictable. This means that, if a higher accuracy is required, the ionization gauge must be calibrated for each gas species. The scaling of the sensitivity with the ionization probability has clear limits, because the scaling factor depends on the energy distribution of the electrons in the ionization space, which is usually not known.

### 3.2  Calibration of total pressure vacuum gauges

The larger NMIs will calibrate vacuum gauges by comparison with the known pressure generated in their fundamental standard [8]. Most vacuum gauges, however, are calibrated by comparison with another calibrated vacuum gauge, called a reference gauge. A reference gauge should have a lower measurement uncertainty than the gauge unit under calibration (UUC), or at least have the same measurement uncertainty. The vacuum system with which the comparison is performed must expose the entrance flange of the UUC and that of the reference gauge to the same density and velocity distribution of calibration gas molecules. The same density and velocity distribution of these molecules means the same pressure at the two locations, but not vice versa.

The vacuum system to be used for a calibration by comparison was standardized by ISO 3567. This standard specifies the requirements for the shape of the calibration chamber, the position of gas inlet and pump outlet, the position of flanges for the gauges, the temperature, and it describes the calibration procedures. Two procedures are common: one where the valve to the pump is closed, and the other where it is open and a stationary equilibrium is established in the calibration chamber.

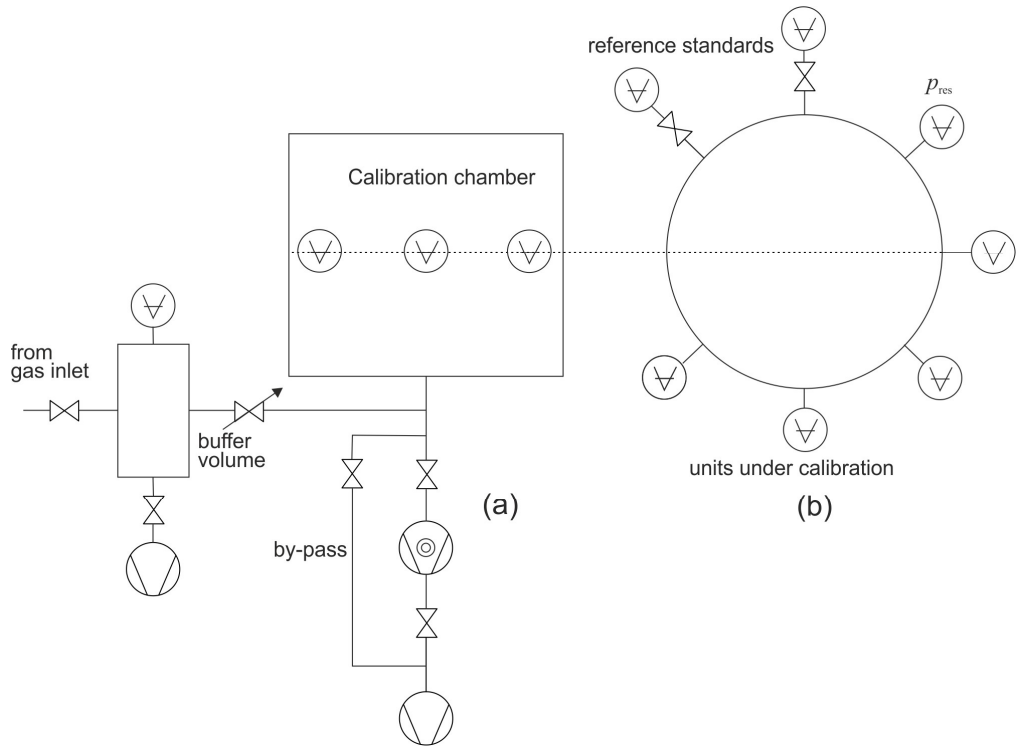

**Fig. 3:** Diagram of a comparison calibration system that realizes the ISO 3567. (a) Side view (b) top view. The calibration chamber has to have cylindrical symmetry (including a sphere as a possibility), gas inlet and outlet must be located on the cylindrical axis, and the flanges to adopt the reference standards and the units to calibrate must lie on the same equatorial plane. The buffer volume helps to stabilize the gas flow in, and a bypass of the turbomolecular pump helps to evacuate the calibration chamber after venting. From Ref. [8].

ISO 3567 also addresses how to write a calibration report and to treat measurement uncertainties due to the calibration system and procedure. The full list of uncertainties and how to evaluate them is given in ISO 27893. The two basic standards ISO 3567 and 27893 are complemented by additional standards which describe the relevant parameters, calibration guidelines, and uncertainties for special types of gauges. One for the Pirani gauges is already published (ISO 19685) while the publication of the standard for capacitance diaphragm gauges is expected to be published in the year 2018 (Table 1). Therein, it is also described how the long-term instability should be determined.

**Table 1:** Overview of ISO standards for total pressure vacuum gauge calibration

| ISO number | Publication date | Title | Content and scope |
|---|---|---|---|
| **3567** | 2011 | Vacuum gauges—calibration by direct comparison with a reference gauge | Basic standard for how a comparison calibration system for vacuum gauges must appear. Procedures how to calibrate vacuum gauges by comparison. Measurement uncertainty. |
| **27893** | 2012 | Vacuum technology—vacuum gauges—evaluation of the uncertainties on results of calibrations by direct comparison with a reference gauge | Guidelines for the determination and reporting of measurement uncertainties arising during vacuum gauge calibration in accordance with ISO 3567. Sum and quotient model of quantity under calibration. List of possible uncertainties. |
| **27894** | 2009 | Vacuum technology—vacuum gauges—specifications for hot cathode ionization gauges | Specifications and terms related to the calibration, use, and measurement uncertainties of ionization gauges with hot emission cathode |
| **19685** | 2017 | Vacuum technology—vacuum gauges—specifications, calibration, and measurement uncertainties for Pirani gauges | Complements ISO 3567 and ISO 27893 when characterizing or calibrating Pirani gauges or using them as reference gauges |
| **20146** | 2018 (expected) | Vacuum technology—vacuum gauges—specifications, calibration, and measurement uncertainties for capacitance diaphragm gauge | Complements ISO 3567 and ISO 27893 when characterizing, calibrating, or using capacitance diaphragm gauges as reference gauges |

Another standard related to vacuum gauges is ISO 27894 'Specifications for hot cathode ionization gauges'. This defines the terms relevant for the calibration of hot cathode ionization gauges and the necessary specifications to be given for a calibration.

For further details of the calibrations of vacuum gauges the reader is referred to Refs. [2][3], and [8] and references therein.

# 4 Partial pressure analysers

The measurement of partial pressures is dominated by quadrupole mass spectrometers (QMS). Compared to magnetic sector spectrometers they deliver a broad measurement range and are convenient to use. Optical methods such as Fourier-transform infrared spectroscopy (FTIR) systems cannot detect all of the gas species important for vacuum technology and are difficult to adapt to vacuum systems. In addition, they are quite expensive, while the price for QMS went down over the past decades. QMS are sometimes no more expensive than an accurate total pressure vacuum gauge. For this reason it makes sense to discuss QMS as the only partial pressure analyser.

QMS are also known under the name residual gas analyser (RGA). This term describes one of the possible functions of the QMS, but not all of them. The QMS is also used to control pressures in processes, to detect endpoints in a sputter event, and so on. For this reason, we use the term QMS.

## 4.1 Elements of a quadrupole mass spectrometer (QMS)

A QMS consists of an ion source, a mass filter, and an ion detector (Fig. 4). Control and data-acquisition electronics provides the necessary DC and AC/RF voltages for these three elements and detects the ion current. It is equipped with a communication interface by which a user can control the QMS with a suitable software normally provided with the QMS.

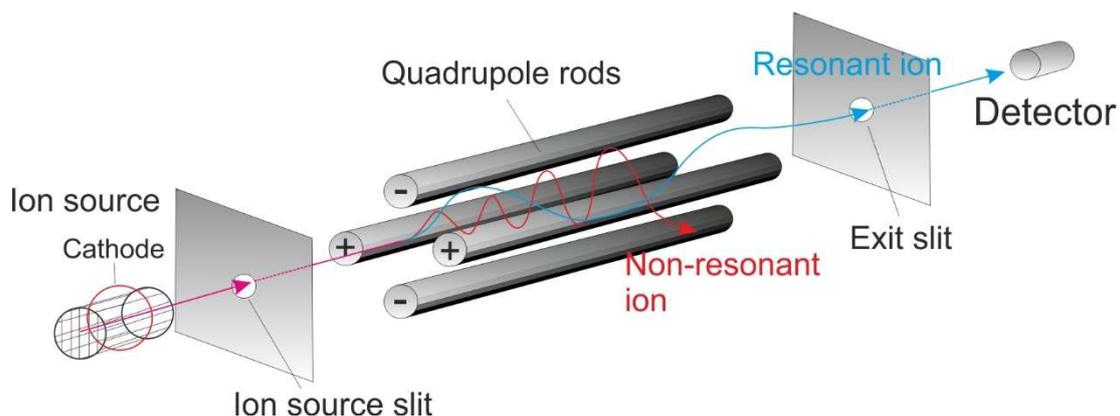

**Fig. 4:** Elements of a quadrupole mass spectrometer

The ion source is very similar to an ionization gauge: electrons emitted from a hot filament are accelerated into an ionization grid space where they ionize neutral molecules by impact. Different types of ion sources such as the open ion source, closed ion source, and molecular beam ion source are common. Open ion sources exhibit a high conductance to the environment of the source and are designed to measure the gas mixture in a vacuum chamber. The gas density in the chamber and in the ion source should be the same. Closed ion sources sample the gas to be analysed and are usually equipped with a differential pumping system that reduces the pressure from the inlet to ion source and the mass filter and detector. A molecular beam ion source is characterized by a collimated beam of process gas focused through the ionization region without colliding with the walls of the ion chamber.

The ions are extracted from the source by a simple or more elaborate electrostatic lens design into the mass filter. The mass filter at any point in time is tuned to a single value of mass-to-charge ratio $m/z$ ($z=1, 2, \ldots$ is the charge state of the ion) and ideally rejects all other values. The ions with the correct ratio can reach the detector at the end of the filter. Scanning through a range of $m/z$ with the filter and recording the ion current as a function of the settings of the mass filter produces a mass spectrum.

The detector, in the simplest case, consists of a Faraday cup, often just a plate, and a current meter. For amplification, secondary electron multipliers (SEM), multichannel plates (MCP), discrete or continuous dynode electron multipliers (DDEM, CDEM) are applied.

## 4.2 The quadrupole mass filter and its theory

The quadrupole mass filter is based on four, carefully parallel aligned, circular rods, aligned on a common alignment circle at 90° apart. The rods are diagonally connected to the same voltage source as shown in **Fig. 5**. The voltage is the sum of the field axis potential *FA* plus a DC voltage *U* and an AC voltage $u \cdot \cos \omega t$ in the radiofrequency range. *FA* is common to both voltage sources while *U* and *u* are of opposite sign. In the centre of the circle an electrical field develops which causes an oscillation of the injected ions perpendicular to the central alignment axis of the rods. Only ions with a suitable value of *m/z* will experience an oscillation of stable amplitude while for other ions the amplitude will build up and the ions will either hit one of the rods or escape out of the quadrupole field between the rods. These ions cannot reach the detector.

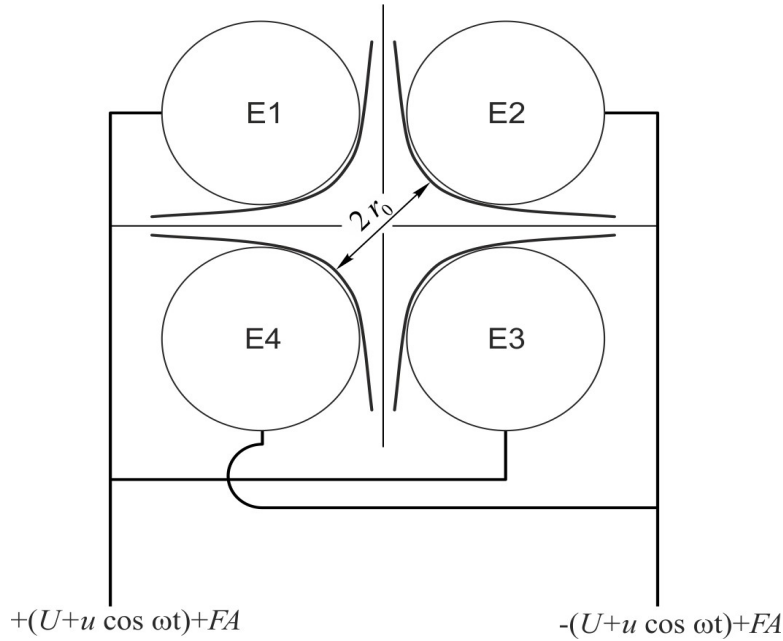

+(*U*+*u* cos ωt)+*FA*        -(*U*+*u* cos ωt)+*FA*

**Fig. 5:** Cross-section of the quadrupole mass filter assembly

The complete problem of the motion equation for ions can be solved as an eigenvalue problem for the two motions in *x* and *y* direction (the *z*-axis is the central alignment axis of the rods). Stable states of the oscillatory motion of an ion in the quadrupole field only exist for a special combination of two parameters, *a* and *q*. These are given by the following equations (note that *z* is the charge state, for $r_0$ see **Fig. 5**, $f = \omega/2\pi$ is the radio frequency [RF], ω the sinusoidal frequency):

$$a = 0.194 \frac{zU}{mr_0^2 f^2},$$

$$q = 0.097 \frac{zu}{mr_0^2 f^2}.$$

(1)

In all other cases, the amplitudes (the *x* and *y* coordinates) of the ions become very large as time, respectively path length along the quadrupole, increases. Since we have a cylindrical symmetry, the problem is symmetric in *x* and *y* and we have either regions where only the *x* coordinate is stable or the *y* coordinate is stable and we have a small area where both *x* and *y* are stable (**Error! Reference source not found.**).

This is the interesting region for quadrupole mass spectrometers and it is enlarged in Fig. 7.

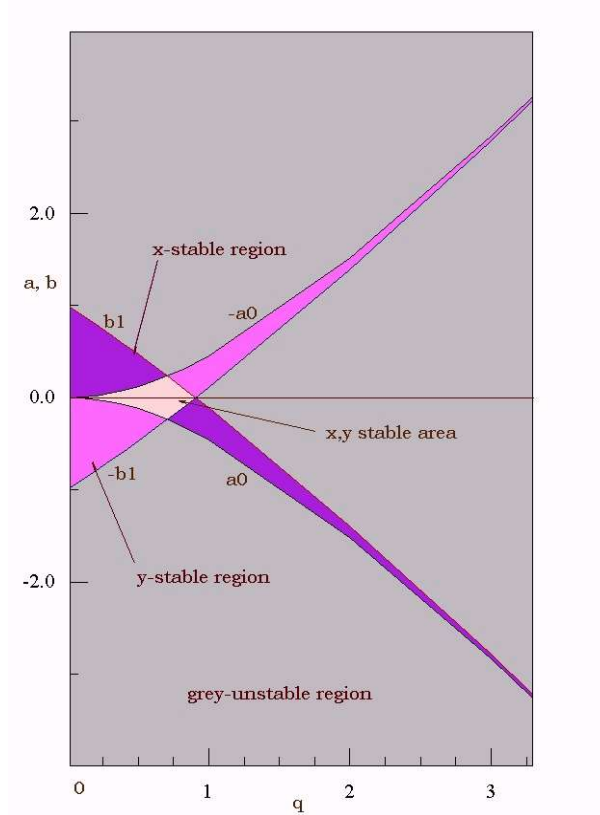

**Fig. 6** The Mathieu diagram: stability regions for the motion in the *x*-direction (*a, q*) and *y*-direction (*b, q*)

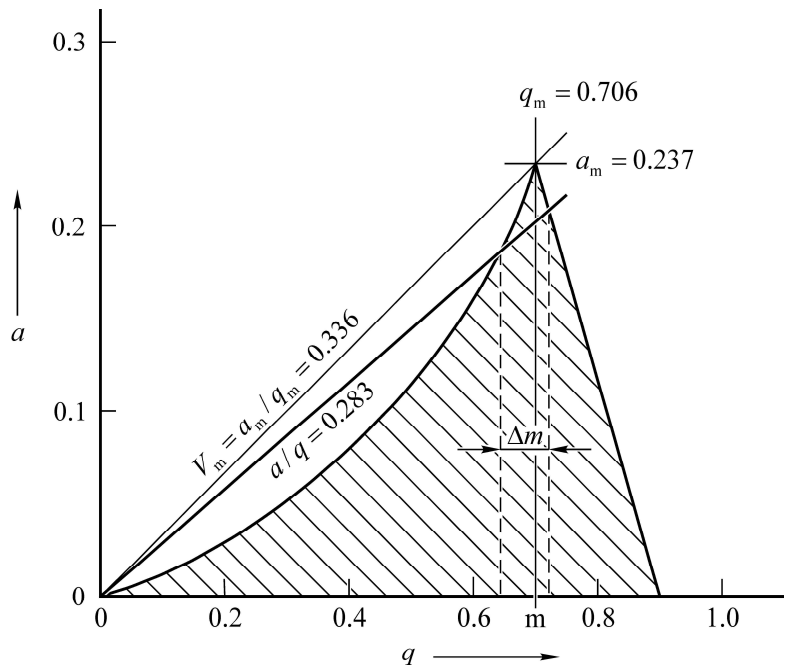

**Fig. 7:** Common stability region for both *x*- and *y*-directions (the upper triangle of Fig.6)

Each ion occupies a certain working point (*a,q*) in the diagram of Fig. 7. A double charged ion (*z* = 2) has the same working point as a single charged ion with half of its mass. Ions of all masses lie on a common straight line, since the ratio *a/q* does not depend on the charge or mass, but solely on the ratio *U/u* of DC to RF voltage:

$$\frac{a}{q} = \frac{2U}{u}. \tag{5}$$

When

$$\frac{2U}{u} < 0.336 \tag{6}$$

a detection of ions is possible. The voltages *u* and *U* define the minimum and maximum masses with a stable trajectory. The max−min difference defines the Δ*m* accepted by the mass filter. Close to the tip of the stability diagram the transmission probability *T* is proportional to the resolving power *R*:

$$T \propto R = \frac{m}{\Delta m} = \frac{1.5064}{1 - \frac{U/u}{(U/u)_{max}}}. \tag{7}$$

This equation shows that a high stability of *U/u* is required. For example, for *R*=100 the voltage ratio is 0.9849, for *R*=110 this ratio is 0.9863. This difference of 0.14% corresponds to a 10% peak width change and shows that a high stability of *U/u* is required for overall stability. The quality of the filtering depends not only on the stability of the voltages, but also on the diameter, the alignment, and the length of the rods. The larger the diameter of the rods the better the approximation to the ideally hyperbolic shape of the rods. The hyperbolic shape would generate the ideal quadrupole field between the rods. With increasing length of the rods, more oscillations and a larger amplitude build-up occur during the travel so that more ions with incorrect values of *m/e* will be taken out by the filter.

To detect higher masses, *U* and *u* must be increased. This has the consequence (**Fig. 8**) that Δ*m* increases with mass numbers (scan line 'W' in Fig. 9). This unwanted effect can be overcome by choosing a scan line with slightly different slope which does not pass through the origin (Fig. 9). The disadvantage of the scan line 'R' with constant resolution is that *m* = 1 (hydrogen ion) cannot be filtered, an effect called the hydrogen blast.

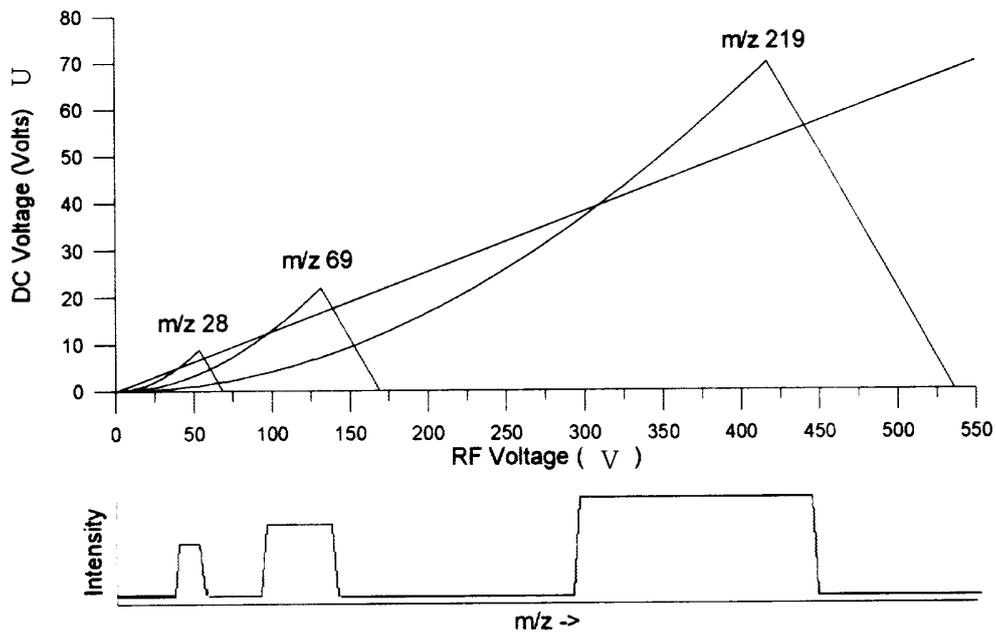

**Fig. 8:** Three stability triangles for different DC and RF voltages on a common straight line through origin

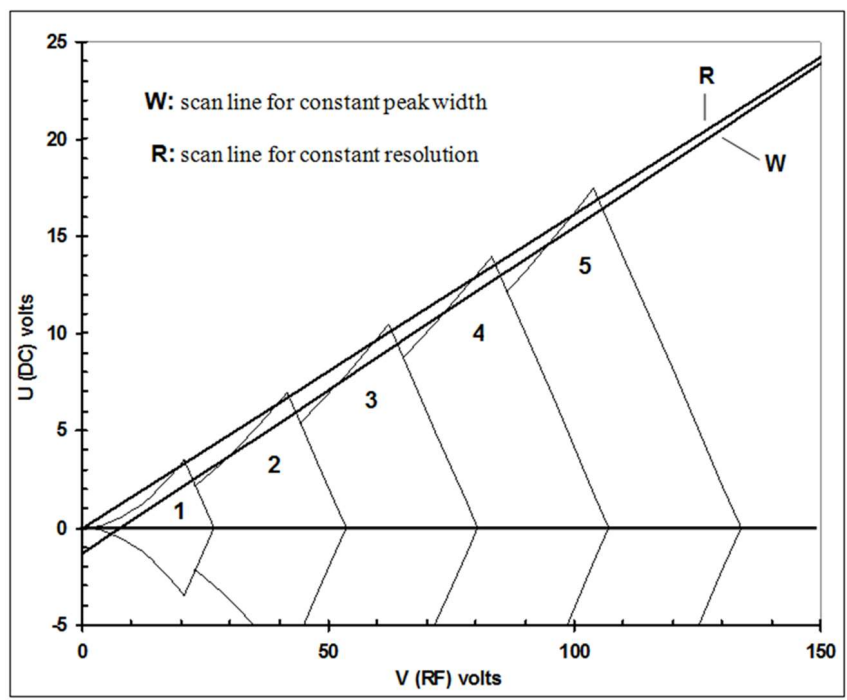

**Fig. 9**: Scan lines for constant peak width (W) and constant resolution (R) along the line

The transmission probability $T$ is approximately given by the ratio of the area above the stability line to the area below it. When $\Delta m$ is kept constant, $(m/\Delta m) \cdot T$ is a constant. This means that $T$ is proportional to $1/m$. Among other effects, such as the extraction efficiency of the ions from the source, the sensitivity of a QMS is given by the product of the ionization probability and transmission probability. This has the interesting effect that the sensitivities for helium and nitrogen in a typical QMS are about the same, since the higher transmission probability 28/4 of helium compared to nitrogen is compensated by the lower ionization probability (**Fig. 10**).

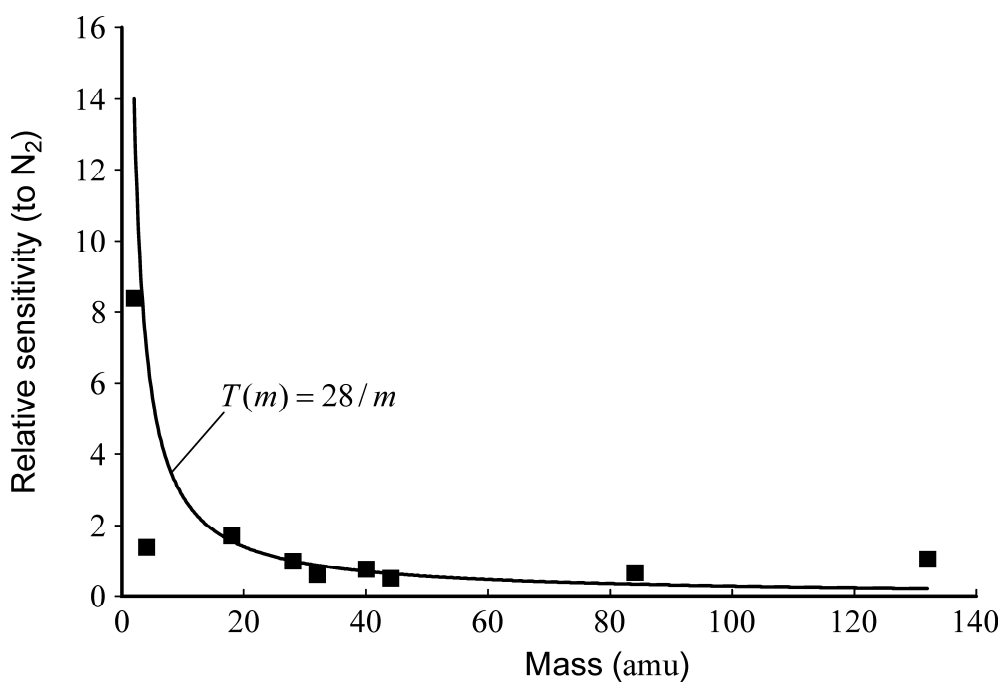

**Fig. 10:** Relative sensitivities and ion transmission for some common gas species in a typical QMS

**4.3 Measurements with quadrupole mass spectrometers**

While the mass scale ($m/z$) is normally not a challenge for the applications in vacuum technology, the most prominent exception is that CO and $N_2$ cannot be distinguished at peak $m/z$ =28 by most QMS, the partial pressure measurement suffers from many problems.

(i) The sensitivity depends greatly on the settings of the instrument: emission current, electron energy, ion energy, resolution, scan speed, scan mode, multiplier gain.
(ii) There are no general rules for the settings. Each type of QMS, and sometimes even individual instruments of the same type, react differently on changes of the settings.
(iii) The sensitivity depends on total pressure.
(iv) The sensitivity for one gas species depends on the presence of another gas species.
(v) The relative sensitivity for a gas species compared to nitrogen cannot be predicted.
(vi) The ion source fragments molecules and generates multiple charged atoms.
(vii) The QMS produces and measures mass peaks generated by itself.
(viii) The reproducibility of the sensitivity of a QMS is poor.

A systematic overview of the metrological characteristics of different types of QMS was given in Ref. [12]. Two highlights are mentioned here: 1) The detected ion current is generally not proportional to the emission current. 2) The helium sensitivity of helium in argon can change by a factor of 7 when the argon pressure is increased from zero to $10^{-2}$ Pa.

In conclusion, the ion current or partial pressure signals, depending on the type of QMS, must be viewed with great caution. Large uncertainties of the measured signals and partial pressures have to be considered.

As a consequence of problems (iii) and (iv), it is not possible to calibrate a QMS in a general manner. Only if the user can define the gas mixture and total pressure range can the QMS be calibrated for this mixture in the given range. On the other hand, due to the poor reproducibility (long-term stability) of the QMS, the uncertainty of a calibration does not need to be very low. Relative uncertainties of 5% to 10% are sufficient.

To make the QMS a more useful instrument, the responsible Technical Committee TC 112 of ISO performed standardization work (**Table 2**). In ISO 14291 terms and definitions for QMS are defined which include important physical parameters of the instrument such as mass resolving power, mass number stability, minimum detectable partial pressure, minimum detectable concentration, fragmentation factor, linear response range, stability of peak height, and so on. ISO 14291 also states which specifications and parameters should be given by a manufacturer.

Table 2: Overview of ISO standards for QMS characterization

| ISO number | Publication date | Title | Content and scope |
|---|---|---|---|
| **14291** | 2012 | Vacuum gauges—definitions and specifications for quadrupole mass spectrometers | Terms and definitions of quadrupole mass spectrometers. Specifications and parameters required for proper calibration and for maintaining the quality of partial pressure measurement. QMSs with an ion source of the electron impact ionization type. $m/z$ typically <300. |
| **TS 20175** | 2018 | Vacuum technology—vacuum gauges—characterization of quadrupole mass spectrometers for partial pressure measurement | Parameters, vacuum systems and procedures to characterize quadrupole mass spectrometers. QMSs with an ion source of the electron impact ionization type. No differential pumping. $m/z$ <300. Adjustment of QMS left to user's experience. |

To make the specifications in the data sheets of different manufacturers comparable, ISO TS 20175 was developed. This Technical Specification was published in April 2018. Therein, it is described how the different parameters, such as minimum detectable partial pressure or minimum detectable concentration, have to be measured. Besides the general characterization of a QMS to determine the specifications in a data sheet, other procedures are described to enable the user of a QMS to make traceable and quantitative measurements of leak rates and outgassing rates. Different vacuum systems and procedures, including the very important in situ measurements for calibration of a QMS, are specified.